\documentclass[useAMS,usenatbib,letterpaper]{mn2e}
 \usepackage{times}
 \usepackage{graphicx}
\usepackage{natbib}

\newcommand\aj{{AJ}}%
\newcommand\apj{{ApJ}}%
\newcommand\apjl{{ApJ}}%
\newcommand\aap{{A\&A}}%
\newcommand\aaps{{A\&AS}}%
\newcommand\mnras{{MNRAS}}%
\newcommand\nat{{Nature}}%
\newcommand\physrep{{Phys.~Rep.}}%
\let\lesssim=\la
\let\gtrsim=\ga

\def\gs{\mathrel{\raise0.35ex\hbox{$\scriptstyle >$}\kern-0.6em
\lower0.40ex\hbox{{$\scriptstyle \sim$}}}}
\def\ls{\mathrel{\raise0.35ex\hbox{$\scriptstyle <$}\kern-0.6em
\lower0.40ex\hbox{{$\scriptstyle \sim$}}}}

\newcommand{\hc}{\citetalias{hick11qsoclust}}

\newcommand{\average}[1]{\ensuremath{\langle#1\rangle} }
\newcommand{\fnm}{\footnotemark}
\newcommand{\fnt}{\footnotetext}

\newcommand{\msun}{M_{\sun}}

\newcommand{\hminus}{$h^{-1}$}
\newcommand{\hmpc}{$h^{-1}$\,Mpc}
\newcommand{\hmsun}{$h^{-1}$\,$M_{\sun}$}

\begin{document}

\voffset=-0.6in

\title[Clustering of SMGs]{The LABOCA Survey of the Extended {\it Chandra} Deep Field South: Clustering of submillimetre galaxies}
\author[Ryan C. Hickox et al.]{Ryan C.\ Hickox$^{\! 1,2,3}$\thanks{E-mail:
ryan.c.hickox@dartmouth.edu}, J.\,L.\ Wardlow$^{\! 1,4}$, Ian Smail$^{\! 5}$, A.\,D.\ Myers$^{\! 6}$, D.\,M.\ Alexander$^{\! 1}$, \newauthor A.\,M.\ Swinbank$^{\! 5}$, A.\,L.\,R.\ Danielson$^{\! 5}$, J.\,P.\,Stott$^{\! 1}$, S.\,C.\ Chapman$^{\! 7}$, K.\,E.\,K.\ Coppin$^{\! 8}$,\newauthor J.\,S.\ Dunlop$^{\! 9}$, E.\ Gawiser$^{\! 10}$, D.\ Lutz$^{\! 11}$, P.\ van der Werf$^{\! 12}$, A.\ Wei{\ss}$^{\! 13}$  \\
$^{1}$Department of Physics, Durham University, South Road, Durham DH1 3LE\\
$^{2}$STFC Postdoctoral Fellow\\
$^{3}$Department of Physics and Astronomy, Dartmouth College, 6127 Wilder Laboratory, Hanover, NH 03755, USA\\
$^{4}$Department of Physics \& Astronomy, University of California, Irvine, CA 92697, USA\\
$^{5}$Institute for Computational Cosmology, Durham University, South Road, Durham DH1 3LE\\
$^{6}$Department of Physics and Astronomy, University of Wyoming, Laramie, WY 82071, USA\\
$^{7}$Institute of Astronomy, Madingley Road, Cambridge CB3 0HA\\
$^{8}$Department of Physics, McGill University, Ernest Rutherford Building, 3600 Rue University, Montreal, Quebec H3A 2T8, Canada\\
$^{9}$Institute for Astronomy, University of Edinburgh, Royal Observatory, Edinburgh EH9 3HJ\\
$^{10}$Department of Physics and Astronomy, Rutgers, The State University of New Jersey, Piscataway, NJ 08854, USA\\
$^{11}$Max-Planck-Institut f\"{u}r extraterrestrische Physik, Postfach 1312, 85741 Garching, Germany\\
$^{12}$Leiden Observatory, Leiden University, NL 2300 RA Leiden, The Netherlands\\
$^{13}$Max-Planck-Institut f\"{u}r Radioastronomie, Auf dem H\"{u}gel 69, 53121, Bonn, Germany\\}


\pagerange{\pageref{firstpage}--\pageref{lastpage}} \pubyear{2011}

\maketitle

\label{firstpage}

\begin{abstract}
  We present a measurement of the spatial clustering of submillimetre
  galaxies (SMGs) at $z=1$--3. Using data from the 870 $\mu$m LABOCA
  submillimetre survey of the Extended {\it Chandra} Deep Field South,
  we employ a novel technique to measure the cross-correlation between
  SMGs and galaxies, accounting for the full probability distributions
  for photometric redshifts of the galaxies.  From the observed
  projected two-point cross-correlation function we derive the linear
  bias and characteristic dark matter halo masses for the SMGs. We
  detect clustering in the cross-correlation between SMGs and galaxies
  at the $>4 \sigma$ level. Accounting for the clustering of galaxies
  from their autocorrelation function, we estimate an autocorrelation
  length for SMGs of $r_0 = 7.7^{+1.8}_{-2.3}$ \hmpc\ assuming a
  power-law slope $\gamma=1.8$, and derive a corresponding dark matter
  halo mass of $\log(M_{\rm halo} [h^{-1}\; M_{\sun}])
  =12.8^{+0.3}_{-0.5}$.  Based on the evolution of dark matter haloes
  derived from simulations, we show that that the $z=0$ descendants of
  SMGs are typically massive ($\sim$\,2--3 $L^*$) elliptical galaxies residing
  in moderate- to high-mass groups ($\log(M_{\rm halo} [h^{-1}\; M_{\sun}]) =13.3^{+0.3}_{-0.5}$). From the observed clustering we
  estimate an SMG lifetime of $\sim$100 Myr, consistent with lifetimes
  derived from gas consumption times and star-formation timescales,
  although with considerable uncertainties. The clustering of SMGs at
  $z\sim2$ is consistent with measurements for optically-selected
  quasi-stellar objects (QSOs), supporting evolutionary scenarios in
  which powerful starbursts and QSOs occur in the same systems. Given
  that SMGs reside in haloes of characteristic mass
  $\sim$\,$6\times10^{12}$ \hmsun, we demonstrate that the redshift
  distribution of SMGs can be described remarkably well by the
  combination of two effects: the cosmological growth of structure and
  the evolution of the molecular gas fraction in galaxies. We conclude
  that the powerful starbursts in SMGs likely represent a short-lived
  but universal phase in massive galaxy evolution, associated with the
  transition between cold gas-rich, star-forming galaxies and
  passively evolving systems.

\end{abstract}


\clearpage
\section{Introduction}
\label{sec.intro}

Submillimetre galaxies (SMGs) are a population of high-redshift
ultraluminous infrared galaxies (ULIRGs) selected through their
redshifted far-infrared emission in the submillimetre waveband
\citep[e.g.,][]{smai97smg,barg98smg,hugh98smg,blai02smg}. The redshift
distribution of this population appears to peak at $z\sim 2.5$
\citep[e.g.,][]{chap03smgz, chap05smg, ward11less}, so that SMGs are
at their commonest around the same epoch as the peak in powerful
active galactic nuclei (AGN) and specifically quasi-stellar objects
(QSOs) \citep[e.g.,][]{rich06qlf, asse11qsolfunc}. This correspondence
may indicate an evolutionary link between SMGs and QSOs, similar to
that suggested at low redshift between ULIRGs and QSOs by
\citet{sand88}. However there is little direct overlap ($\sim$\,a few
percent) between the high-redshift SMG and QSO populations
\citep[e.g.,][]{page04submm, chap05smg, stev05submm, alex08bhmass,
  ward11less}.  The immense far-infrared luminosities of SMGs are
widely believed to arise from intense, but highly-obscured, gas-rich
starbursts \citep[e.g.,][]{grev05smgco, alex05, pope08smgspitz,
  tacc06smgco, tacc08smg, ivis11smgco}, suggesting that they may
represent the formation phase of the most massive local galaxies:
giant ellipticals \citep[e.g.,][]{eale99smg,swin06smg}.


SMGs and QSOs may thus represent phases in an evolutionary sequence
that eventually results in the population of local massive elliptical
galaxies. This is a compelling picture, but testing the evolutionary
links is challenging due to the lack of an easily-measured and
conserved observable to tie the various populations together. For
example, the stellar masses of both QSOs and SMGs are difficult to
measure reliably due to either the brightness of the nuclear emission
in the QSOs \citep[e.g.,][]{croo04qsohost,kota09qsohost} or strong
dust obscuration and potentially complex star-formation histories for
the SMGs (e.g., \citealt{hain11smgmass, ward11less}; but see also
\citealt{dunl11smg, mich11smg}), while the details of the
high-redshift star formation that produced local massive elliptical
galaxies are likewise poorly constrained
\citep[e.g.,][]{alla09ellage}. Deriving dynamical masses for QSO hosts
from rest-frame optical spectroscopy is difficult due to the very
broad emission lines from the AGN, while dynamical mass measurements
using CO emission in gas-rich QSOs are also challenging, due to the
potential non-isotropic orientation of the QSO hosts on the sky and
the lack of high-resolution velocity fields necessary to solve for
this \citep{copp08qsoco}, as well as the general difficulties in
modeling CO kinematics \citep[e.g.,][]{tacc06smgco, both10smgco,
  enge10smgco}.

%
%
\begin{figure}
    \includegraphics[width=\columnwidth]{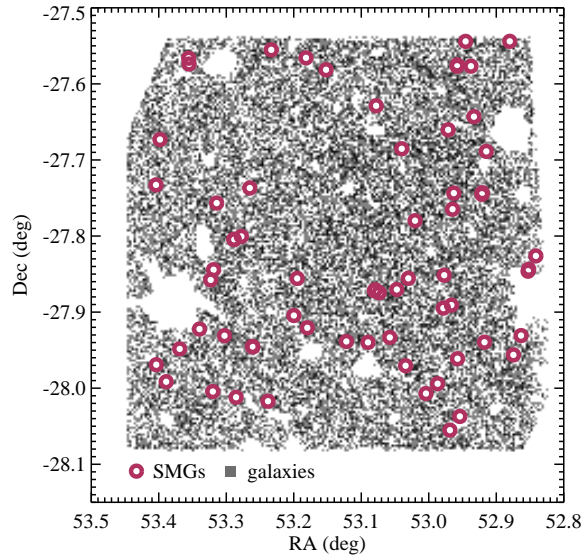}
    \caption{Two-dimensional distribution of the 50 LESS SMGs and
      $\sim$\,50,000 IRAC galaxies in the ECDFS that are used in our
      analysis. The SMGs shown represent the subset of the 126 SMGs in
      the full LESS sample \citep{weis09less} that are in the redshift
      range $1 < z < 3$ and are in regions of good photometry, and so
      are used in this analysis. The IRAC galaxies are chosen to
      reside at $0.5 < z < 3.5$. The SMGs are shown here individually,
      while the density of galaxies is given by the grayscale. The
      blank areas represent regions which are excluded from the
      analysis, including areas of poor photometry (for example around
      bright stars) or additional sources identified by eye in the
      vicinity of SMG, as discussed in \S\ref{sec.samples}. The high
      density of IRAC galaxies in the field enables an accurate
      measurement of the SMG-galaxy cross-correlation
      function.\label{fig.sky}}
\end{figure}

%
%
\begin{figure}
   \includegraphics[width=\columnwidth]{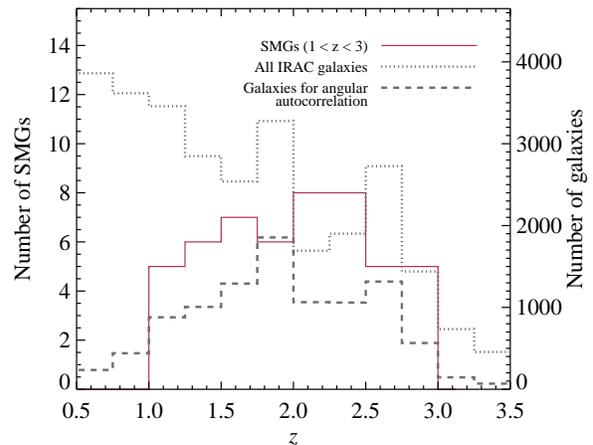}
\caption{Redshift distributions for the IRAC galaxy sample in the
  redshift range $0.5< z < 3.5$ (dotted line), and the SMG sample in
  the range $1 < z < 3$ (solid line). The histogram for galaxies has
  been scaled so that the distribution can be directly compared to
  that of the SMGs. Also shown is the redshift distribution for 11,241
  galaxies (dashed line) selected to match the overlap in the redshift
  distributions of the SMGs and galaxies, as used in the galaxy
  autocorrelation measurement (\S\ref{sec.galauto}).  For the SMGs,
  44\% have spectroscopic redshifts, while the remainder of the SMGs
  and all the IRAC galaxies have redshift estimates from photometric
  redshift calculations \citep{ward11less}.
  \label{fig.z}}
\end{figure}

Another possibility is to compare source populations via the masses of
their central black holes. For QSOs and the population of SMGs that
contain broad-line AGN, the black hole mass can be estimated using
virial techniques based on the broad emission lines
\citep[e.g.,][]{vest02bhmass, pete04bhmass, vest06, koll06,
  shen08bhmass}. Such studies generally find that SMGs have small
black holes relative to the local black hole-galaxy mass relations
\citep[e.g.,][]{alex08bhmass, carr11qsogroup}, while the black holes
in $z\sim 2$ QSOs tend to lie above the local relation, with masses
similar to those in local massive ellipticals
\citep[e.g.,][]{deca10mbhevol, benn10bhmass, merl10bhmass}. These
results suggest that SMGs represent an earlier evolutionary stage,
prior to the QSO phase in which the black hole reaches its final
mass. However, high-redshift virial black hole mass estimates are
highly uncertain \citep[e.g.,][]{marc08radpress, fine10bhmass,
  netz10bhmass} and may suffer from significant selection effects
\citep[e.g.,][]{laue07mbhbias, shen10mbhbias, kell10qsoedd}, and so
conclusions about connections between populations are necessarily
limited.

The difficulties discussed above lead us to take another route to
compare SMGs to high-redshift QSOs and low-redshift ellipticals:
through their clustering.  Spatial correlation measurements provide
information about the characteristic bias and hence mass of the haloes
in which galaxies reside \citep[e.g.,][]{kais84clust, bard86gauss},
and so provide a robust mass estimate that is free of many of the
systematics in measuring stellar or black hole masses. The observed
clustering of SMGs and QSOs can thus allow us to test whether these
populations are found in similar haloes and so may evolve into each
other over short timescales.  With knowledge of how haloes evolve over
cosmic time \citep[e.g.,][]{lace93merge, fakh10halorate}, we can also
explore the links to modern elliptical galaxies
\citep[e.g.,][]{over03radio}, as well as the higher-redshift
progenitors of SMGs. Clustering measurements can also provide
constraints on theoretical studies that explore the nature of SMGs in
a cosmological context. Recent models for SMGs as relatively long-lived
($>$\,0.5\,Gyr) star formation episodes in the most massive galaxies,
driven by the early collapse of the dark matter halo
\citep{xia11smgclust}, or powered by steady accretion of intergalactic
gas \citep{dave10smg}, yield strong clustering for bright sources (850 $\mu$m
fluxes $>$\,a few mJy) with correlation lengths $r_0
\gtrsim\,10$ \hmpc. In contrast, models in which SMGs are short-lived
bursts in less massive galaxies, with large luminosities produced by a
top-heavy initial mass function, predict significantly weaker
clustering with $r_0 \sim 6$ \hmpc\ \citep{alme11smgclust}.

Attempts to measure the clustering of SMGs from their projected
two-dimensional distribution on the sky have for the most part been
ambiguous \citep{scot02scuba, bory03scuba, webb03submmlbg, weis09less,
  will11smgclust, lind11mm}.  \citet{weis09less} used the
largest, contiguous extragalactic 870-$\mu$m survey (of the Extended
{\it Chandra} Deep Field South; ECDFS), to derive the clustering of
$\gs 5$-mJy SMGs from their projected distribution on the sky.  They
estimated a correlation length of $13\pm 6 h^{-1}$\,Mpc.  Most
recently, \citet{will11smgclust} analysed a 1100-$\mu$m survey of a
region of the COSMOS field and placed 1-$\sigma$ upper limits on the
clustering of bright SMGs (with apparent 870-$\mu$m fluxes $\gs
8$--10\,mJy) of $\gs 6$--12\,$h^{-1}$\,Mpc. 

Other work has attempted to improve on angular correlation
measurements by including redshift information. Using the
spectroscopic redshift survey of 73 SMGs with 870-$\mu$m fluxes of
$\gs 5$\,mJy spread across seven fields from \citet{chap05smg},
\citet{blai04smgclust} estimated a clustering amplitude from the
numbers of pairs of SMGs within a 1000-km\,s$^{-1}$ wide velocity
window.  They derived an effective correlation length of $6.9\pm 2.1$
\hmpc, suggesting that SMGs are strongly clustered. However
their methodology was subsequently criticised by \citep{adel05pairs},
who suggested that accounting for angular clustering of sources and
the redshift selection function significantly increases the
uncertainties. Using data from the {\em Chandra} Deep
Field-North, \citet{blak06smgclust} computed the angular
cross-correlation between SMGs and galaxies in slices of spectroscopic
and photometric redshift. They obtained a significant SMG-galaxy
cross-correlation signal, with hints that SMGs are more strongly
clustered than the optically-selected galaxies, although with only
marginal ($\sim$\,2$\sigma$) significance. Previous work has therefore
pointed toward SMGs being a strongly clustered population, but their
precise clustering amplitude, along with their relationship to QSOs
and ellipticals, remains uncertain.

To make improved measurements of the clustering of SMGs, we need
either much larger survey areas (see \citealt{coor10herclust} for a
wide-field clustering measurement for far-IR detected sources) or the
inclusion of redshift information (to allow us to reduce the effects
of projection on our clustering measurements).  To this end, we have
reanalysed the \citet{weis09less} survey of ECDFS using new
spectroscopic and photometric redshift constraints on the counterparts
to SMGs \citep{ward11less} as well as a large catalogue of ``normal''
(less-active) galaxies in the same field. We employ a new clustering
analysis methodology \citep{myer09clust} to calculate the projected
spatial cross-correlation between SMGs and galaxies, to obtain the
tightest constraint to date on the clustering amplitude of SMGs.

This paper is organised as follows. In \S~2 we introduce the SMG and
galaxy samples, and in \S~3 we give an overview of the methodology
used to measure correlation functions and estimate dark matter (DM) halo
masses. In \S~4 we present the results, explore the effects of
photometric redshift errors, compare with previous measurements, and
discuss our results in the context of the physical drivers, lifetimes,
and evolutionary paths of SMGs. In \S~5 we summarise our
conclusions. Throughout this paper we assume a cosmology with
$\Omega_{\rm m}=0.3$ and $\Omega_{\Lambda}=0.7$.  For direct
comparison with other works, we assume $H_0=70$ km s$^{-1}$ Mpc$^{-1}$
(except for comoving distances and DM halo masses, which are
explicitly given in terms of $h=H_0/(100$ km s$^{-1}$ Mpc$^{-1})$).
In order to easily compare to estimated halo masses in other recent
works on QSO clustering \citep[e.g.,][]{croo05, myer06clust,
  daan08clust, ross09qsoclust}, we assume a normalisation for the
matter power spectrum of $\sigma_8 = 0.84$.  All quoted uncertainties
are $1\sigma$ (68\% confidence).

\section[]{SMG and galaxy samples}
\label{sec.samples}

Our SMG sample comes from the survey of the ECDFS using the Large APEX
BOlometer CAmera \citep[LABOCA]{siri09laboca} on the Atacama
Pathfinder EXperiment \citep[APEX]{gust06apex} 12-m telescope (the
LABOCA ECDFS Submillimetre Survey, or LESS;
\citealt{weis09less}). LESS mapped the full 0.35 deg$^2$ ECDFS to a
870-$\mu$m noise level of $\sim 1.2$ mJy beam$^{-1}$ and detected 126
SMGs at $> 3.7 \sigma$ significance \citep[equivalent to a
  false-detection rate of $\sim 4$\%]{weis09less}. Radio and
mid-infrared counterparts to LESS SMGs were identified by
\citet{bigg11less} using a maximum-likelihood technique. Spectroscopic
and photometric redshifts were obtained for a significant fraction of
these counterparts by \citet{ward11less} and we refer the reader to
that work for more details. For this study, we restrict our analysis
to the 50 SMGs that have secure counterparts at $z=1$--3 and do not
lie close to bright stars (as discussed below). The upper limit of
$z=3$ on the sample is included to maximize overlap in redshift space
with the galaxy sample, in order to obtain a significant
cross-correlation signal, while the lower bound of $z=1$ is included
to prevent the SMG sample from being biased toward low redshifts. Of
the SMGs in the sample, 22 SMGs (44\%) have spectroscopic redshifts
(Danielson et al., in preparation) and the remainder have photometric
redshifts with a typical precision of $\sigma_z/(1+z) \sim 0.1$
\citep{ward11less}.  The 870-$\mu$m flux distribution for the SMGs
having secure counterparts \citep{bigg11less} is consistent with that
for all LESS SMGs \citet{weis09less}, indicating that the requirement
that SMGs have secure counterparts does not strongly bias the fluxes
of our SMG sample.

\begin{figure}
   \includegraphics[width=\columnwidth]{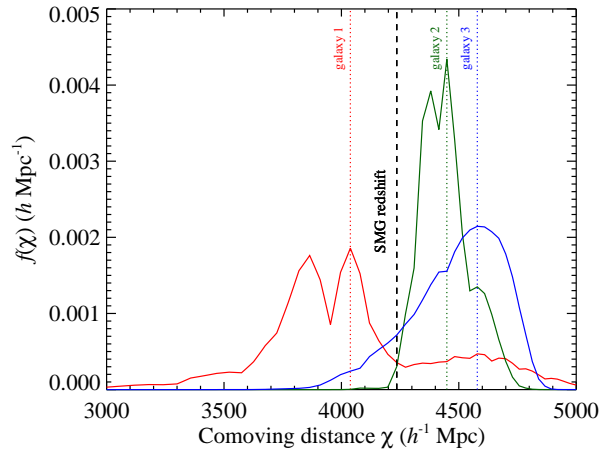}
\caption{Example probability distribution functions for three IRAC galaxies and an SMG.
We mark the  ``best'' (peak) comoving distance for each
  galaxy. Note that for each galaxy in this example, the line-of-sight
  distance between the ``peak'' redshift of the galaxy and the SMG
  redshift is far too large for them to be physically associated.
  However, because of the uncertainty in the galaxy redshifts (shown
  by the PDFs), there is a non-negligible probability that the
  galaxies lie close to the line-of-sight distance of the SMG.  
\label{fig.pdf}
}
\end{figure}

For the cross-correlation analysis, we also require a comparison
population in the same field.  For this we adopt the $\sim$\,50,000
galaxies detected in the Spitzer IRAC/MUSYC Public Legacy Survey in
the Extended CDF-South \citep{dame11simple}. We use an IRAC selected
sample to ensure that each galaxy has photometry in a sufficient
number of bands, and over a wide enough wavelength range, to allow
robust estimates of photometric redshift. Photo-$z$s are calculated
using template fits to the optical and IRAC photometry in an identical
method to that used for the SMGs (see \citealt{ward11less}). The fits
are performed with {\sc hyper-z} \citep{bolz00} and the resulting
redshift distribution, compared to that for the SMGs, is shown in
Figure~\ref{fig.z}.  The photometric analysis uses chi-squared minimisation,
which allows the calculation of confidence intervals for the best-fit
redshift. These can be presented as a probability distribution
function (PDF) for the redshift, or equivalently, the comoving
line-of-sight distance $\chi$ (calculated for our assumed cosmology).
We define the PDF for each galaxy as $f(\chi)$, where $\int f(\chi)
d\chi = 1$. Examples of the PDFs for the galaxies are shown in
Figure~\ref{fig.pdf}.

Finally, in order to calculate the correlation functions, we first
create random catalogues of ``galaxies'' at random positions within
the actual spatial coverage of our survey.  Like many fields, the
ECDFS contains several bright stars with large haloes, around which few
galaxies are detected. Therefore, we use the background map produced
by {\sc SExtractor} \citep{bert96} from the combined IRAC image during
the source extraction procedure to create a mask. This mask is applied
to the random catalogues, the SMGs and the IRAC galaxies, so that the
positions of the random galaxies are unbiased with respect to the SMG
and IRAC galaxy samples, and thus the mask does not affect the
cross-correlation measurement.  As discussed in \citet{bigg11less} and
\citet{ward11less}, some of the SMG identifications were performed
manually by examining the regions around the SMGs. These additional
sources are excluded from the clustering analysis so as not to bias
the results. The sky positions of the SMGs and galaxies that are
outside the masked regions are shown in Figure~\ref{fig.sky}.

%
%

\section{Correlation analysis}

\defcitealias{hick11qsoclust}{H11}
\defcitealias{myer09clust}{M09}

\label{sec.corranal}

To measure the spatial clustering of SMGs, we can in principle derive
the autocorrelation of the SMGs themselves.  However, as we have
discussed, current SMG samples are too limited in size and available
redshift information to make this feasible. Alternatively, we can
measure the {\it cross}-correlation of a population with a sample of
other sources (for example, less-active galaxies) which populate the
same volume \citep[e.g.,][]{gawi01dla, adel05qsoclust, blak06smgclust,
  coil07a, hick09corr}.  The much larger number of galaxies in the
ECDFS ($\sim $\,1000\,$\times$ more than the SMGs in a comparable
redshift range) allows far greater statistical accuracy in the
measurement of clustering.

To calculate the real-space projected cross-correlation function
$w_p(R)$ between SMGs and galaxies we employ a method derived by
\citet{myer09clust}. This method enables us to take advantage of the
full photo-$z$ PDF for each galaxy, by weighting pairs of SMGs and
galaxies based on the probability of their overlap in redshift
space. This method allows us to calculate the SMG-galaxy
cross-correlation using the full sample of $z\approx50,000$ IRAC
galaxies, while the derive the clustering of the galaxies themselves
using a smaller sample that is selected to match the overlap in the
redshift distributions of the galaxies and SMGs. Our clustering
analysis is identical in most respects to the QSO-galaxy
cross-correlation study presented in \citet[][hereafter
  H11]{hick11qsoclust}. Because the method is somewhat involved, we
present only the key details here and refer the reader to \hc\ for a
full discussion.

\subsection{Cross-correlation method}
\label{sec.crosscorr}

The two-point correlation function $\xi(r)$ is defined as the
probability above Poisson of finding a galaxy in a volume element $dV$
at a physical separation $r$ from another randomly chosen galaxy, such
that
\begin{equation}
dP=n[1+\xi(r)]dV,
\end{equation}
where $n$ is the mean space density of the galaxies in the sample.
The projected correlation function $w_p(R)$ is defined as the integral
of  $\xi(r)$ along the line
of sight,  
\begin{equation}
\label{eqn.wpint}
w_p(R)=2\int_{0}^{\pi_{\rm max}}\xi(R,\pi)d\pi,
\end{equation}
where $R$ and $\pi$ are the projected comoving separations between
galaxies in the directions perpendicular and parallel, respectively,
to the mean line of sight from the observer to the two galaxies.  By
integrating along the line of sight, we eliminate redshift-space
distortions owing to the peculiar motions of galaxies, which distort
the line-of-sight distances measured from redshifts.  $w_p(R)$ has
been used to measure correlations in a number of surveys \citep[e.g.,][]{zeha05a, li06agnclust, gill07c,coil07a, coil08galclust, wake08radio, myer09clust, hick09corr,  coil09xclust, gill09xclust, krum10xclust, dono10clust,  hick11qsoclust, star11xclust, alle11xclust}.


In the range of separations $0.3\lesssim r \lesssim50$ \hminus\ Mpc,
$\xi(r)$ for galaxies and QSOs is roughly observed to be a
power-law,
\begin{equation}
\xi(r)=(r/r_0)^{-\gamma},
\label{eqn.plaw}
\end{equation}
with $\gamma$ typically $\approx$1.8 \citep[e.g.,][]{zeha05a,
  coil08galclust, coil07a, ross09qsoclust}.  For sufficiently large
$\pi_{\rm max}$ such that we average over all line-of-sight peculiar
velocities, $w_p(R)$ can be directly related to $\xi(r)$ (for a power
law parameterisation) by
\begin{equation}
w_p(R)=R\left (\frac{r_0}{R}\right)^\gamma
\frac{\Gamma(1/2)\Gamma[(\gamma-1)/2]}{\Gamma(\gamma/2)}.
\label{eqn.wpplaw}
\end{equation}

To calculate $w_p(R)$ for the cross-correlation between SMGs and
galaxies, we use the method of \citetalias{myer09clust}, which
accounts for the photometric redshift probability distribution for
each galaxy individually. Following \citetalias{myer09clust}, the
projected cross-correlation function can be calculated using:

\begin{equation}
w_p(R) = N_R N_S \sum_{i,j} c_{i,j} \frac{D_S D_G (R)}{D_S R_G (R)} - \sum_{i,j} c_{i,j}
\label{eqn.wp}
\end{equation}
where
\begin{equation}
\label{eqn.cij}
c_{i,j} = f_{i,j}\large{/}\sum_{i,j}f_{i,j}^2.
\end{equation}
Here $R$ is the projected comoving distance from each SMG, for a given
angular separation $\theta$ and radial comoving distance to the SMG of
$\chi_*$, such that $R=\chi_* \theta$. $D_S D_G$ and $D_S R_G$ are the
number of SMG--galaxy and SMG--random pairs in each bin of $R$, and
$N_S$ and $N_R$ are the total numbers of SMGs and
random galaxies, respectively. $f_{i,j}$ is defined as the average value
of the radial PDF $f({\chi})$ for each galaxy $i$, in a window of size
$\Delta \chi$ around the comoving distance to each spectroscopic
source $j$.  We use $\Delta \chi = 100$ \hmpc\ to effectively
eliminate redshift space distortions, although the results are
insensitive to the details of this choice. We refer the reader to
\citetalias{myer09clust} and \hc\ for a detailed derivation and
discussion of these equations. In this calculation as well as in the
galaxy autocorrelation, we account for the integral constraint as
described in \hc. This correction increases the observed clustering
amplitude by $\approx$15\%.

\subsection{Galaxy autocorrelation}

\label{sec.galauto}
To estimate DM halo masses for the SMGs, we calculate the
relative bias between SMGs and galaxies, from which we derive the
absolute bias of the SMGs relative to DM.  As discussed
below, calculation of absolute bias (and thus halo mass) requires a
measurement of the autocorrelation function of the IRAC galaxies.  The
large size of the galaxy sample enables us to derive the clustering of
the galaxies accurately from the angular autocorrelation function
$\omega(\theta)$ alone.  Although we expect the photometric redshifts
for the IRAC galaxies to be reasonably well-constrained (as discussed
in \S~\ref{sec.samples}), by using the angular correlation function we
minimize any uncertainties relating to individual galaxy photo-$z$s
for this part of the analysis. The resulting clustering measured for
the galaxies has significantly smaller uncertainties than that for the
SMG-galaxy cross-correlation.

We calculate the angular autocorrelation function $\omega(\theta)$ using the \citet{land93} estimator:
\begin{equation}
\omega(\theta)=\frac{1}{RR}(DD-2DR+RR),
\label{eqn.xidef1a}
\end{equation}
where $DD$, $DR$, and $RR$ are the number of data-data, data-random,
and random-random galaxy pairs, respectively, at a separation
$\theta$, where each term is scaled according to the total numbers of
SMGs, galaxies, and randoms.  

The galaxy autocorrelation varies with redshift, owing to the
evolution of large scale structure, and because the use of a
flux-limited sample means we select more luminous galaxies at higher
$z$.  This will affect the measurements of relative bias between SMGs
and galaxies, since the redshift distribution of the SMGs peaks at
higher $z$ than that for the galaxies and so relatively higher-$z$
galaxies dominate the cross-correlation signal.  To account for this
in our measurement of galaxy autocorrelation, we randomly select
galaxies based on the overlap of the PDFs with the SMGs in comoving
distance (in the formalism of \S~\ref{sec.crosscorr} this is $f_{i,j}$
for each galaxy, averaged all SMGs).  We select the galaxies so their
distribution in redshift is equivalent to the {\em weighted}
distribution for all galaxies (weighted by $\average{f_{i,j}}$).  The
redshift distribution of this galaxy sample is shown in
Figure~\ref{fig.z}.  We use this smaller galaxy sample to calculate
the angular autocorrelation of IRAC galaxies.

\subsection{Uncertainties and model fits}

We estimate uncertainties on the clustering directly from the data
using bootstrap resampling. Following \hc, we divide the field into a
small number of sub-areas (we
choose $N_{\rm sub} = 8$), and for each bootstrap sample we randomly
draw a total of $3 N_{\rm sub}$ sub-areas (with replacement), which
has been shown to best approximate the intrinsic uncertainties in the
clustering amplitude \citep{norb09clust}. To account for shot noise
owing to the relatively small size of the SMG sample, we take the sets of
$3 N_{\rm sub}$ bootstrap sub-areas and randomly draw from them (with
replacement) a sample of sources (SMGs or galaxies) equal in size to
the parent sample; only pairs including these sources are used in the
resulting cross-correlation calculation. We use the bootstrap results
to derive the covariance between different bins of $R$, calculating
the covariance matrix using Equation~12 of \hc.

We fit the observed $w_p(R)$ with two models: a power law and a simple
bias model (described in \S~\ref{sec.absbias}). We compute model
parameters by minimising $\chi^2$ (taking into account the covariance
matrix as in Equation~13 of \hc) and derive 1$\sigma$ errors in
each parameter by the range for which $\Delta \chi^2 = 1$. We use the
same formalism for computing fits to the angular correlation
functions, where $\omega(\theta) = A\theta^{-\delta}.$ We convert $A$
and $\delta$ to real-space clustering parameters $r_0$ and $\gamma$
following the procedure described in \S~4.6 of \hc.

\subsection{Absolute bias and dark matter halo mass}
\label{sec.absbias}
The masses of the DM haloes in which galaxies and SMGs reside
are reflected in their absolute clustering bias $b_{\rm abs}$ relative
to the DM distribution. The linear bias $b_{\rm abs}^2$ is
given by the ratio of the autocorrelation function of the galaxies (or
SMGs) to that of the DM. We determine $b_{\rm abs}$ following
the method outlined in \S~4.7 of \hc, similar to the approach used
previously by a number of studies \citep[e.g.,][]{myer06clust,
  myer07clust1, coil07a, coil08galclust, coil09xclust, hick09corr}; in
what follows we briefly describe this procedure.

We first calculate the two-point autocorrelation of DM as a
function of redshift. We use the {\sc halofit} code of
\citet{smit03dm} assuming our standard cosmology, and the slope of the
initial fluctuation power spectrum, $\Gamma=\Omega_m h=0.21$, to
derive the DM power spectrum, and thus its projected
correlation function $w_p^{\rm DM}(R)$, averaged over the redshift
distribution for which the SMGs and galaxies overlap. We then fit the
observed $w_p(R)$ of the SMG-galaxy cross-correlation, on scales
$0.3$--15 \hmpc, with a model comprising a simple linear scaling of
$w_p^{\rm DM}(R)$.  The best-fit linear scaling of the DM
correlation function corresponds to $b_S b_G$, the product of the
linear biases for the SMGs and galaxies, respectively. This simple
model produces a goodness-of-fit comparable to that of the power-law
model in which the slope $\gamma$ is allowed to float.

To determine $b_S$ we therefore need to estimate $b_G$.  We obtain
$b_G$ for the galaxies from their angular autocorrelation in a similar
manner to that applied to the SMG--galaxy cross-correlation. Again we
calculate the autocorrelation for the DM $\omega_{DM} (\theta)$, by
integrating the power spectrum from {\sc halofit} using Equation~(A6)
of \citet{myer07clust1}. We fit the observed $\omega(\theta)$ with a
linear scaling of $\omega_{DM} (\theta)$ on scales
$0.3$\arcmin--10$\arcmin$ (corresponding to 0.3--10 \hmpc\ at
$z=2$). This linear scaling corresponds to $b_G^2$ and thus (combined
with the cross-correlation measurement) yields the SMG bias
$b_S$. Finally, we convert $b_G$ and $b_S$ to $M_{\rm halo}$ using the
prescription of \citet{shet01halo}, as described in \hc. This
characteristic $M_{\rm halo}$ corresponds to the top-hat virial mass
\citep[see e.g.,][and references therein]{peeb93book}, in the
simplified case in which all objects in a given sample reside in
haloes of the same mass. This assumption is justified by the fact (as
discussed below in \S~\ref{sec.lifetime}) that SMGs have a very small
number density compared to the population of similarly-clustered DM
haloes, such that it is reasonable that SMGs may occupy haloes in a
relatively narrow range in mass. We note that this method differs from
some prescriptions in the literature which assume that sources occupy
all haloes above some minimum mass; this is particularly relevant for
populations with high number densities that could exceed the numbers
of available DM haloes over a limited mass range. Given the halo mass
function at $z\sim2$ \citep[e.g.,][]{tink08halo} the derived minimum
mass is typically a factor of $\sim$2 lower, for the same clustering
amplitude, than the ``average'' mass quoted here.

%
%
\begin{figure}
    \includegraphics[width=\columnwidth]{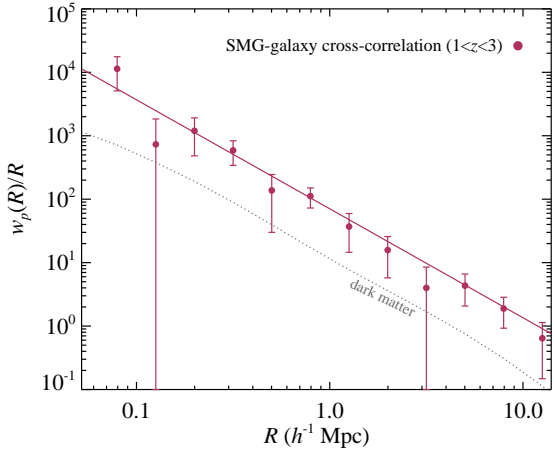}
    \caption{The projected SMG-galaxy cross-correlation function
      (derived using Equation~\ref{eqn.wp}).  Uncertainties are
      estimated from bootstrap resampling.  A power-law fit to
      $w_p(R)$ is shown by the solid line, and the projected
      correlation function for DM is shown by the dotted line.  Fits are
      performed over the range in separation of $R=$\,0.3--15
      \hmpc. Both the power law model with $\gamma = 1.8$ and a linear
      scaling of the DM correlation function provide
      satisfactory fits to the observed $w_p(R)$.  Together with the
      observed galaxy autocorrelation, this measurement
      yields the clustering amplitude and DM halo mass for the SMGs,
      as described in \S~\ref{sec.results}. \label{fig.corr}}
\end{figure}

\section{Results and Discussion}
\label{sec.results}

%
%
\begin{table*} \caption{Correlation results}
\label{tbl.corr}

\begin{minipage}{14cm}
\begin{tabular}{lccccccccc}
\hline
 &
 &
 &
\multicolumn{3}{c}{Power law fit$^{c}$} &
\multicolumn{3}{c}{Bias model fit$^d$} &
Halo mass$^e$ \\
Subset &
${N_{\rm src}}^{a}$ &
$\average{z}^{b}$ &
$r_0$ ($h^{-1}$ Mpc) &
$\gamma$ &
$\chi^2_\nu$ &
$b_S b_G$ ($b_G^2$) &
$b_S$ ($b_G$) &
$\chi^2_\nu$ &
($\log{h^{-1}\; M_{\sun}}$) \\
\hline
SMGs & 50 & 2.02 & $7.7^{+1.8}_{-2.3}$ & $1.8\pm0.2$ & 0.8 & $5.83\pm1.36$ & $3.37\pm0.82$ & 0.7 & $12.8^{+ 0.3}_{- 0.5}$ \\
galaxies & 11,241 & 2.13 & $3.3\pm0.3$ & $1.8\pm0.2$ & 1.8 & $2.99\pm0.40$ & $1.73\pm0.12$ & 1.8 & $11.5\pm 0.2$ \\
\hline
\end{tabular}

\begin{tabular}{p{14cm}}
  $^a$ Number of objects in the SMG sample and in the galaxy sample
  used for the galaxy autocorrelation. \\ $^b$ Median redshift for the
  SMG sample and for the galaxy sample used for the galaxy
  autocorrelation. \\ $^c$ Power law model parameters are for the
  autocorrelation of SMGs (derived from SMG-galaxy projected
  spatial cross-correlation, along with the galaxy angular
  autocorrelation) and galaxies (derived from their angular
  autocorrelation). \\ $^d$ Parameters derived from the observed
  linear fit of the DM model to the observed correlation
  function, in order to obtain the the absolute bias
  for the SMGs and galaxies (denoted $b_S$ and $b_G$,
  respectively). The linear scaling from the fit corresponds to $b_S
  b_G$ for the SMG-galaxy cross-correlation, and $b_G^2$ for the
  galaxy autocorrelation, which in turn yield $b_G$ and $b_S$. \\ $^e$
  DM halo mass derived from the absolute bias, using the method
  described in \S~\ref{sec.absbias}. \\
\end{tabular}
\end{minipage}

\end{table*}

The projected cross-correlation function of the SMG sample with the
IRAC galaxies is shown in Figure~\ref{fig.corr}.  We plot the best-fit
power-law model, and show the correlation function of the DM
calculated as in \S~\ref{sec.absbias}, which we fit to the data
through a linear scaling. The power-law and linear bias fit parameters
are presented in in Table~\ref{tbl.corr}.  For SMGs the observed
real-space projected cross-correlation is well-detected on all scales
from 0.1--15 \hmpc, and the power-law fits return $\gamma\sim 1.8$,
similar to many previous correlation function measurements for
galaxies \citep[e.g.,][]{zeha05a, coil08galclust} and QSOs
\citep[e.g.,][]{coil07a, ross09qsoclust}.  The best-fit parameters for
the SMG-galaxy cross-correlation are $r_{0,SG}=5.3\pm0.8$ \hmpc,
$\gamma=1.7\pm0.2$. If we fix the value of $\gamma$ to 1.8, we obtain
$r_{0,SG} = 5.1\pm 0.6$ \hmpc, corresponding to a clustering signal that is significant at the $>$\,4$\sigma$ level, the most significant
measurement of SMG clustering to date. From the fit of the DM model,
we obtain $b_S b_G = 5.83\pm1.36$.

\begin{figure}
    \includegraphics[width=\columnwidth]{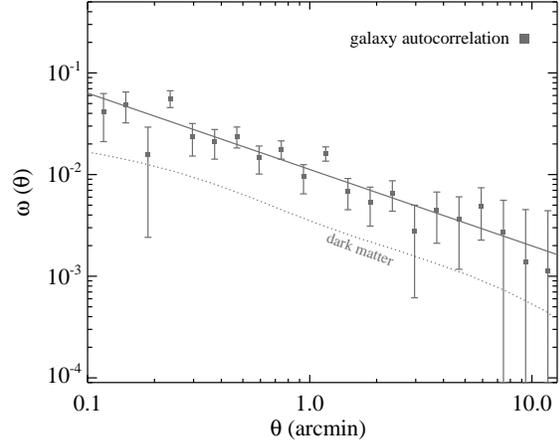}
  \caption{The angular autocorrelation function of IRAC galaxies,
    selected to match the overlap of the SMGs and galaxies in redshift
    space.  Uncertainties are estimated from bootstrap resampling.
    The angular correlation function for DM, evaluated for
    the redshift distributions of the galaxies, is shown by the dotted
    gray line.  The power law fit was performed on scales
    0.3\arcmin--10\arcmin\ and is shown as the solid line. Both the power law model with $\delta = 0.8$ and a linear
      scaling of the DM correlation function provide
      satisfactory fits to the observed $\omega(\theta)$. The observed amplitude of the galaxy autocorrelation yields the absolute bias of the galaxies, which we use to obtain the absolute bias and DM halo mass of the SMGs.
      \label{fig.galang}}
\end{figure}

We next compute the autocorrelation of IRAC galaxies for the sample
described in \S~\ref{sec.galauto}.  The observed $\omega(\theta)$ is shown
in Fig.~\ref{fig.galang}, along with the corresponding power-law fit
and scaled correlation function for DM, calculated as
discussed in \S\ref{sec.absbias}. Fit parameters are given in
Table~\ref{tbl.corr}. The power-law model fits well on the chosen
scales of 0.3\arcmin--10\arcmin.  The best-fit power law parameters
are $r_{0,GG} = 3.3 \pm 0.3$ and $\gamma = 1.8\pm0.2$, and the best-fit
scaled DM model yields $b_G^2 = 2.99\pm 0.40$ or $b_G = 1.73\pm
0.12$.  

This accurate value for $b_G$ yields $b_S=3.37\pm 0.82$ for the SMGs.
Converting this to DM halo mass using the prescription of
\citet{shet01halo} as described in \S\ref{sec.absbias}, we arrive at
$\log{(M_{\rm halo} [h^{-1}\; M_{\sun}])}= 12.8^{+0.3}_{-0.5}$. The
corresponding halo mass for the galaxies is $\log{(M_{\rm halo} [h^{-1}\;
  M_{\sun}])}= 11.5\pm0.2$.

%
%

For comparison with other studies that attempted to directly measure
the autocorrelation function of SMG, it is useful to present the SMG
clustering in terms of effective power-law parameters for their
autocorrelation. Assuming linear bias, the SMG autocorrelation can be
inferred from the cross-correlation by $\xi_{SS} = \xi_{SG}^2 /
\xi_{GG}$ \citep[e.g.,][]{coil09xclust}.  Adopting a fixed $\gamma=1.8$ for
the SMG-galaxy cross-correlation, we thus obtain $r_{0,SS} =
7.7^{+1.8}_{-2.3}$ \hmpc\ for the autocorrelation of the SMGs.

\subsection{Effects of SMG photo-$z$ errors}

One uncertainty in our estimate of $w_p(R)$ for the SMG-galaxy
cross-correlation is due to the lack of accurate (that is,
spectroscopic) redshifts for roughly half of the SMG population. As
described in \S~\ref{sec.corranal}, in calculating $w_p(R)$ for the
cross-correlation, we simply assume that the SMGs lie exactly at the
best redshifts from the photo-$z$ analysis of \citet{ward11less}.  Any
uncertainties in the SMGs photo-$z$s could therefore affect the
resulting clustering measurement. (Note that photo-$z$ uncertainties
in the galaxies are accounted for implicitly in the correlation
analysis, as we utilize the full galaxy photo-$z$ PDFs.)  To examine
the effects of SMG photo-$z$ errors, we follow the procedure outlined
in \S~6.3 of \hc. We take advantage of the 44\% of SMGs that do have
spectroscopic redshifts, and determine how errors in those redshifts
affect the observed correlation amplitude.

Specifically, we shift the redshifts of the spectroscopic SMGs by
offsets $\Delta z / (1+z)$ selected from a Gaussian random
distribution with dispersion $\sigma_z/(1+z)$. To ensure that this
step does not artificially smear out the redshift distribution beyond
the range probed by the galaxies, we require that the random redshifts
lie between $1 < z < 3$; any random redshift that lies outside this
range is discarded and a new redshift is selected from the random
distribution.  Using these new redshifts we recalculate $w_p(R)$,
using the full formalism described in \S~\ref{sec.corranal}.  We
perform the calculation 10 times for each of several values of
$\sigma_z/(1+z)$ from 0.05 up to 0.3 (corresponding to the range of
photo-$z$ uncertainties). For each trial we obtain the relative bias
by calculating the mean ratio of $w_p(R)$, on scales 1--10 \hmpc,
relative to the $w_p(R)$ for the best estimates of redshift.  We then
average the ten trials at each $\sigma_z$, and find that at most the
photo-$z$ errors cause the clustering amplitude to decrease by
$\sim$\,10\%. The precise magnitude of this effect is unclear given
the range of uncertainties in the SMG photo-$z$ estimates, but it is
is significantly smaller than the statistical uncertainties. We
therefore neglect this effect in our final error estimates.


\subsection{Comparison with previous results}

\begin{figure}
  \includegraphics[width=\columnwidth]{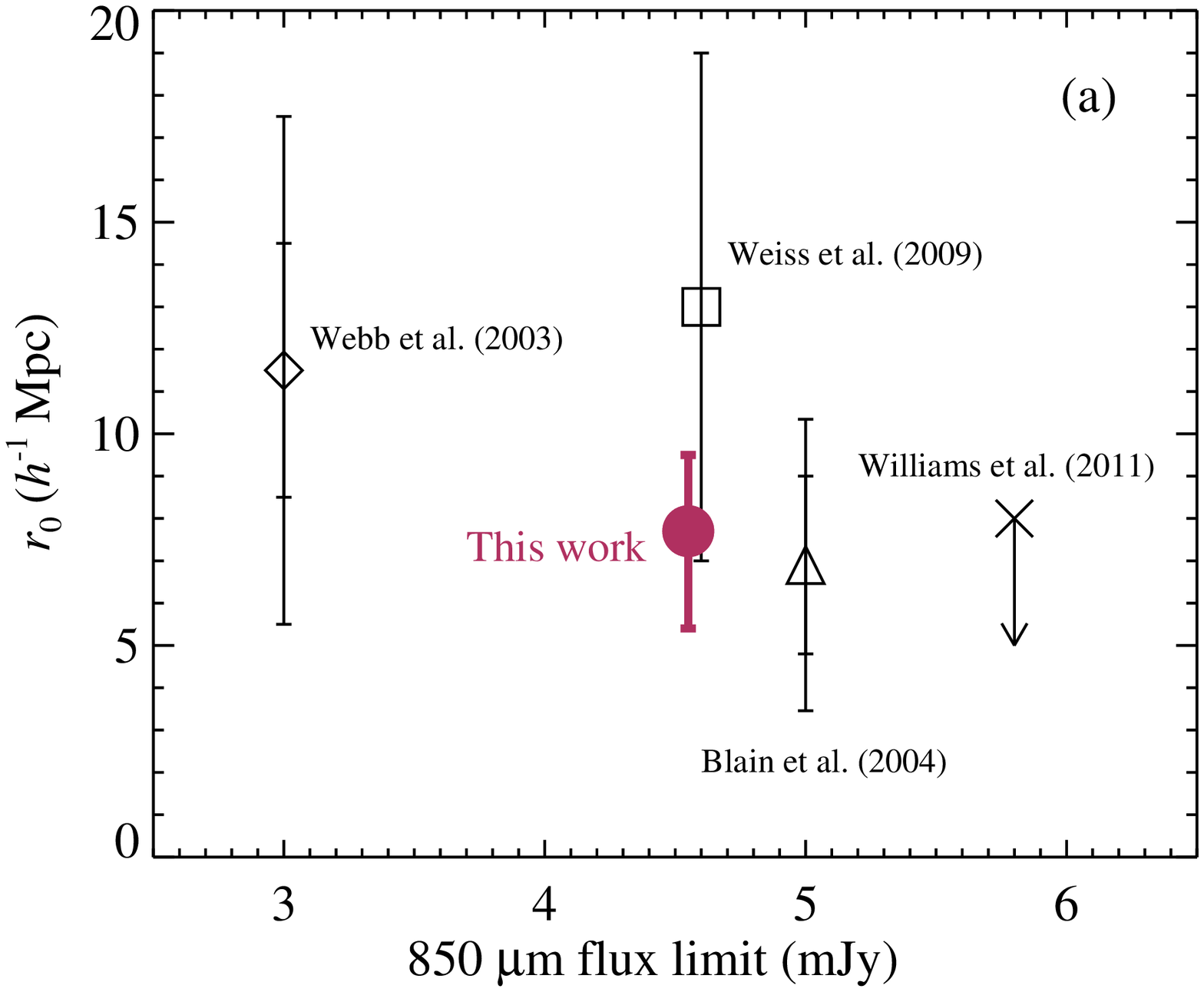}
  \includegraphics[width=\columnwidth]{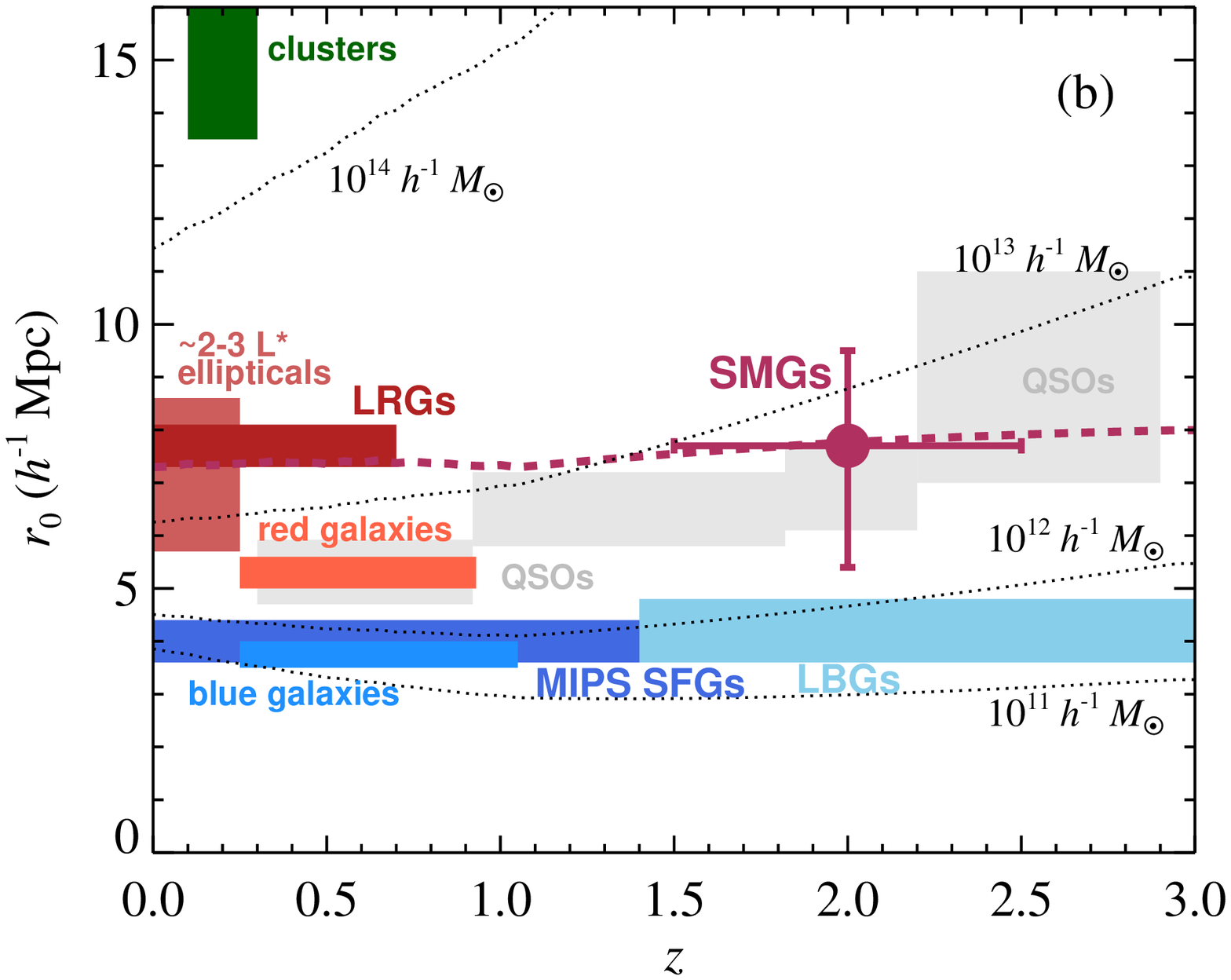}
  \caption{({\em a}) Our new measurement of the autocorrelation length
    $r_0$ for SMGs, compared to previous results using samples with
    similar $\sim$850 $\mu$m flux limits. The two sets of error bars
    on the \citet{webb03submmlbg} measurement indicate statistical
    ($\pm 3$ \hmpc) and systematic ($\pm 3$ \hmpc) uncertainties
    separately. On the \citet{blai04smgclust} measurement, the smaller
    errors represent the uncertainties quoted by the authors, while
    the larger errors account for angular clustering and redshift
    spikes as estimated by \citet{adel05pairs}. Our results are
    consistent with previous measurements and represent a significant
    improvement in precision. ({\em b}) Our measurement of the
    autocorrelation length $r_0$ of SMGs, compared to the approximate
    $r_0$ (with associated measurement uncertainties) for a variety of galaxy and AGN
    populations: optically-selected SDSS QSOs at $0 < z < 3$
    \citep{myer06clust, ross09qsoclust}, Lyman-break galaxies (LBGs)
    at $1.5 \lesssim z \lesssim 3.5$ \citep{adel05lbgclust}, MIPS 24
    \micron-selected star-forming galaxies at $0 < z < 1.4$
    \citep{gill07c}, typical red and blue galaxies at $0.25 \lesssim z
    \lesssim 1$ from the AGES \citep{hick09corr} and DEEP2
    \citep{coil08galclust} spectroscopic surveys, luminous red
    galaxies (LRGs) at $0 < z < 0.7$ \citep{wake08lrgclust}, and
    optically-selected galaxy clusters at $0.1 < z < 0.3$
    \citep{estr09maxbcgclust}.  In addition, we show the full range of $r_0$ for low-redshift galaxies with $r$-band luminosities in the range 1.5 to 3.5 $L^*$,  derived from the luminosity dependence of clustering presented by \citet{zeha11galclust}; these luminous galaxies are primarily ellipticals, as discussed in \S~\ref{sec.progen}. Dotted lines show $r_0$ versus redshift
    for DM haloes of different masses. The thick solid line
    shows the expected evolution in $r_0$, accounting for the increase
    in mass of the halo, for a halo with mass corresponding to the
    best-fit estimate for SMGs at $z=2$. The results indicate that
    SMGs are clustered similarly to QSOs at $z\sim2$ and can be
    expected to evolve into luminous elliptical galaxies in the local
    Universe.
    \label{fig.r0}}
\end{figure}

Here we compare our results to other measurements of SMG clustering in
the literature.  The observed clustering may depend on the flux limit
of the submm sample, as discussed by \citet{will11smgclust};
measurements of $r_0$ that use SMG samples with similar submm flux
limits are shown in Figure~\ref{fig.r0}{\em a}. Our measurement is
significantly more accurate than previous measurements, owing to the
inclusion of redshift information and the improved statistics in the
cross-correlation. The uncertainties are comparable to those quoted by
\citet{blai04smgclust} who estimated $r_0$ using counts of close pairs
in redshift space from spectroscopic surveys. However, these authors
did not account for significant additional sources of error, as
discussed by \citet{adel05pairs}. Uncertainties in the redshift
selection function for spectroscopic objects, along with the presence
of redshift spikes and angular clustering of sources, can strongly
impact the number of expected pair counts for an unclustered
distribution, and therefore significantly affect the results for the
clustering amplitude \citep{adel05pairs}. In Figure~\ref{fig.r0}{\em
  a} the large error bars for the \citet{blai04smgclust} point
represent the increase in the uncertainty by 60\% due to angular
clustering of sources and redshift spikes (as estimated by
\citealt{adel05pairs}), but does not include the additional
uncertainty on the redshift selection function. Nonetheless, our
measurement of $r_0$ is consistent with most previous angular
clustering estimates as well as the \citet{blai04smgclust} result, and
represents a significant improvement in precision.

As discussed in \S~\ref{sec.absbias}, we convert the observed
clustering amplitude to $M_{\rm halo}$ by assuming that SMGs obey
simple linear bias relative to the dark matter and reside in haloes of
similar mass.  Motivated by the presence of a large overdensity of
SMGs and powerful star-forming galaxies in one redshift survey field,
\citet{chap09smgspike} proposed that SMGs obey ``complex bias'' that
depends on large-scale environment and merger history, and that they
may reside in somewhat smaller haloes than would be inferred from a
linear bias model. Future studies using significantly larger SMG
samples may be able to confirm the existence of more complex
clustering, but for the present analysis we adopt the simplest
scenario and derive $M_{\rm halo}$ assuming linear bias.

The characteristic halo mass we measure for SMGs is similar to that
measured for bright far-IR sources (with fluxes $>30$ mJy at 250
$\mu$m) detected by the {\em Herschel Space Observatory} using an
angular clustering analysis \citep{coor10herclust}. While it remains
uncertain to what extent bright 250 $\mu$m sources and 850
$\mu$m-selected SMGs represent a common population, both samples
comprise the luminous end of the star-forming galaxy population
detected at those wavelengths and so may represent physically similar
systems. In contrast, our observed SMG clustering is significantly
stronger than that reported by \citet{ambl11herclust} for
``submillimetre galaxies'' based on a power-spectrum analysis of {\em
  Herschel} 350 $\mu$m maps, which yields a minimum $M_{\rm halo}$ of
$\sim$\,$3\times10^{11}$ $\msun$. The differences in clustering
amplitude compared to SMGs result from the fact that the power
spectrum analysis includes unresolved faint sources corresponding to
far fainter far-IR luminosities, characteristic of typical $z\sim2$
star-forming galaxies rather than the powerful, luminous starbursts
that are conventionally referred to as SMGs in the literature.

%
%

\subsection{Progenitors and descendants of SMGs}

\label{sec.progen}

Our improved clustering measurement allows us to place SMGs in the
context of the cosmological history of star formation and growth of DM
structures. Because the clustering amplitude of dark
matter haloes and their evolution with redshift are directly predicted by
simulations and analytic theory, we can use the observed clustering to
connect the SMG populations to their descendants and progenitors, estimate lifetimes,
and constrain  starburst triggering mechanisms.

We first compare the clustering amplitude of SMGs with other galaxy
populations over a range of redshifts\fnm.  \fnt{\citet{myer06clust}
  and \citet{ross09qsoclust} determine $r_0$ from QSOs assuming a
  power law correlation function with $\gamma = 2$. To estimate $r_0$
  for $\gamma=1.8$, we multiply the quoted values by 0.8, appropriate
  for fits over the range $1 \lesssim R \lesssim 100$ \hmpc.}
Figure~\ref{fig.r0}{\em b} shows the approximate ranges of
measurements of $r_0$ for a variety of galaxy and AGN populations. We
also show the evolution of $r_0$ with redshift for DM haloes
of different masses, determined by fitting a power law with
$\gamma=1.8$ to the DM correlation function output by {\sc
  halofit}. Finally, we show the observed $r_0$ for the current SMG
sample, along with the expected evolution in $r_0$ for haloes that have
the observed $M_{\rm halo}$ for SMGs at $z=2$, calculated using the
median growth rate of haloes as a function of $M_{\rm halo}$ and $z$
\citep{fakh10halorate}\fnm.

Figure \ref{fig.r0}{\em b} shows that while the DM halo mass for the
SMGs will increase with time from $z\sim 2$ to $z=0$, the observed
$r_0$ stays essentially constant, meaning that the progenitors and
descendants of SMGs will be populations with similar clustering
amplitudes. Our measurement of $r_0$ shows that the clustering of SMGs
is consistent with optically-selected QSOs \citep[e.g.,][]{croo05,
  myer06clust, daan08clust, ross09qsoclust}. SMGs are more strongly
clustered than the typical star-forming galaxy populations at all
redshifts \citep[e.g.][]{adel05lbgclust, gill07c, hick09corr,
  zeha11galclust}, and are clustered similarly or weaker than massive,
passive systems \citep[e.g.,][]{quad07drgclust, quad08eroclust,
  wake08lrgclust, blan08bzk, kim11eroclust, zeha11galclust}. The
clustering results indicate that SMGs will likely evolve into the most
massive, luminous early type galaxies at low redshift. We note that
the descendants of typical SMGs are not likely to reside in massive
clusters at $z=0$, but into moderate- to high-mass groups of $\sim$\,a
few $\times10^{13}$ \hmsun. Although {\em some} SMGs could evolve into
massive cluster galaxies, the observed clustering suggests that most
will end up in less massive systems.

\begin{figure}
  \includegraphics[width=\columnwidth]{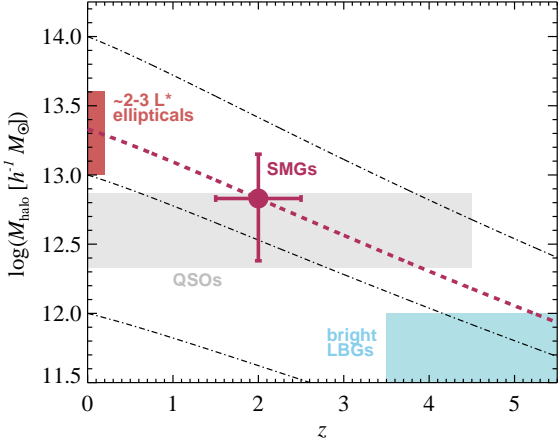}
  \caption{ Broad schematic for the evolution of halo mass versus
    redshift for SMGs, showing the approximate halo masses
    corresponding to likely progenitors and descendants of SMGs. Lines
    indicate the median growth rates of haloes with redshift
    \citep{fakh10halorate}. SMG host haloes are similar to those those
    of QSOs at $z\sim2$, and correspond to bright LBGs at $z\sim 5$
    \citep{hama04lbgclust, lee06lbgclust} and $\sim$\,2--$3 L^*$ ellipticals
    at $z=0$ \citep{zeha11galclust, stot11xcs_submit}.
\label{fig.mhalo}}
\end{figure}

A schematic picture of the evolution of SMGs is shown in Figure
\ref{fig.mhalo}, which shows evolution in the mass of haloes with redshift
as traced by their median growth rate \citep{fakh10halorate}. The
typical progenitors of SMGs would have $M_{\rm halo} \sim 10^{12}$
\hmsun\ at $z\sim5$, which corresponds to the host haloes of bright
LBGs at those redshifts \citep[e.g.,][]{hama04lbgclust,
  lee06lbgclust}. At low redshift, the SMG descendants will have
$M_{\rm halo}=(0.6$--$5)\times 10^{13}$ \hmsun. Halo occupation
distribution fits to galaxy clustering suggest that these haloes host
galaxies with luminosities $L \sim 2$--3$L^*$ \citep{zeha11galclust},
a population dominated by ellipticals with predominantly slow-rotating
kinematics \citep[e.g.,][]{temp11sdssgal, capp11atlas3d}. Assuming
typical mass-to-light ratios for massive galaxies
\citep[e.g.,][]{bald08mass}, these luminosities correspond to stellar
masses $\sim$\,$(1.5$--$2.5)\times10^{11} \msun$, in close agreement
with direct measurements of the relationship between halo mass and
central galaxy stellar mass for X-ray selected groups and clusters,
for which $\log{M_\star} \approx 0.27\log{M_{\rm halo}} + 7.6$
\citep{stot11xcs_submit}.

\fnt{Note that here we use the median growth rate of haloes, which for
  haloes of $\sim10^{13}$ \hmsun\ is $\approx$35\% lower than the {\em
    mean} growth rate, owing to the long high-mass tail in the halo
  mass distribution.}

%
%

\subsection{SMG lifetime and star formation history}

\label{sec.lifetime}

We next estimate the SMG lifetime, making the simple assumption 
that every dark matter halo of similar mass passes through an SMG 
phase\fnm, so that
\begin{equation}
t_{\rm SMG} = \Delta t \frac{n_{\rm SMG}}{n_{\rm halo}},
\label{eqn.lifetime}
\end{equation}
where $\Delta t$ is the time interval over the redshift range covered
by the SMG sample, and $n_{\rm SMG}$ and $n_{\rm halo}$ are the space
densities of SMGs and DM haloes, respectively. 
\fnt{If the
average halo experiences more or fewer SMG phases in the given time
interval, the lifetime of each episode will be correspondingly shorter
or longer, respectively.}

Using the halo mass function of \citet{tink08halo}, the space density
of haloes with $\log{(M_{\rm halo} [h^{-1}\; M_{\sun}])}=
12.8^{+0.3}_{-0.5}$ is ${\rm d}n_{\rm halo}/{\rm d}\ln{M} =
(2.1^{+7.3}_{-1.5})\times10^{-4}$ Mpc$^{-3}$. We adopt a space density
of SMGs at $z\sim 2$ of $\sim$\,$2\times10^{-5}$ Mpc$^{-3}$,
corresponding to results from previous surveys (e.g.,
\citealt{chap05smg, copp06shades}; Schael et al.\ in
preparation). This density is $\sim$\,50\% higher than that observed
in the LESS field \citep{ward11less}, which has been shown to contain
a somewhat smaller density of SMGs compared to other surveys
\citep{weis09less}.

The ratio of these space densities yields a duty cycle (the fraction
of haloes that host an SMG at any given time) of $\sim$\,10\%. We
assume the SMGs occupy the redshift range $1.5 < z < 2.5$, which
includes roughly half of the SMGs in the \citet{ward11less} sample and
corresponds to $\Delta t= 1.6$ Gyr. We thus obtain a lifetime for SMGs
of $t_{\rm SMG} = 110^{+280}_{-80}$ Myr. Clearly, even our improved
measurement of SMG clustering yields only a weak constraint on the
lifetime, but this is consistent with lifetimes estimated from gas
consumption times and star-formation timescales
\citep[e.g.,][]{grev05smgco, tacc06smgco, hain11smgmass} and
theoretical models of SMG fueling through mergers
\citep[e.g.,][]{miho94merge, spri05, nara10smg}.

Constraints on SMG descendants from clustering can also yield insights
into their their formation histories. Measurements of the stellar plus
molecular gas masses of SMGs from SED fitting and dynamical studies
are in the range $\sim$\,$(1$--$5)\times10^{11}$ $\msun$
\citep{swin06smg, ward11less, hain11smgmass, ivis11smgco,
  mich11smg}. While these estimates
can be uncertain by factors of a few, they are in a similar range to
the stellar masses of SMG descendants as indicated by their
clustering, as discussed above. This correspondence suggests that if a
significant fraction of the molecular gas is converted to stars during
the SMG phase, then these galaxies will subsequently experience
relatively little growth in mass from $z\sim 2$ to the present. This
in turn puts limits on the star formation history. Star-forming
galaxies at $z\sim2$ typically exhibit specific star formation rates
of $\dot{M_\star}/M_{\star} \sim 2$ Gyr$^{-1}$ \citep{elba11ms}, at which 
the SMGs would only need to form stars for 500 Myr in order to double
in mass. We may therefore conclude, from the clustering and stellar
masses alone, that the SMGs evolve from star-forming to passive states
relatively quickly (within a Gyr or so) after the starburst phase, and
that the descendants spend most of their remaining time as relatively
passive systems. This scenario is consistent with measurements of the
stellar populations in $\sim$\,2--3 $L^*$ ellipticals, which have typical ages
of $\sim$10 Gyr and show little evidence for younger components
\citep[e.g.,][]{nela05ellage, alla09ellage}, implying that the vast majority of
stars were formed above $z\sim 2$ with little additional star
formation at lower redshifts.

The halo masses of SMGs may also provide insight into the processes
that prevent their descendants from forming new stars. Star formation
can be shut off rapidly at the end of the SMG phase, either by
exhaustion of the gas supply, or by energy input from a QSO
\citep[e.g.,][]{dima05qso, spri05}.  Powerful winds are observed in
luminous AGN \citep[e.g.,][]{feru10outflow,fisc10mrk231, stur11qsoco,
  gree11obsqso} and have also been seen in some SMGs (e.g.,
\citealt{alex10outflow}, Harrison et al.\ in preparation), although
for the SMGs is unclear whether the winds are driven by the starburst
or AGN. Even if the formation of stars is rapidly quenched, over
longer timescales the galaxy would be expected to accrete further gas
from the surrounding halo, resulting in significant additional star
formation \citep[e.g.,][]{bowe06gal, crot06}. Recent work suggests
that energy from accreting supermassive black holes, primarily in the
form of radio-bright relativistic jets, can couple to the hot gas in
the surrounding halo, producing a feedback cycle that prevents rapid
cooling \citep[e.g.,][]{raff08feedback}. This mechanical black hole
feedback is an key ingredient of successful models for the passive
galaxy population \citep[e.g.,][]{crot06, bowe06gal, bowe08flip,
  some08bhev}. Interestingly, the clustering of radio galaxies at
$z\lesssim 0.8$ indicates that they reside in haloes of mass
$\gtrsim$\,$10^{13}$ \hmsun\ \citep[e.g.,][]{wake08radio, hick09corr,
  mand09agnclust, dono10clust, fine11radio}, precisely the
environments that will host the descendants of SMGs. Thus the strong
observed clustering for SMGs can relate them directly to the
radio-bright active galactic nucleus population that may regulate
their subsequent star formation.

\subsection{Evolutionary links with QSOs and the SMG redshift distribution}

\label{sec.evol}

%
%
\begin{figure}
  \includegraphics[width=\columnwidth]{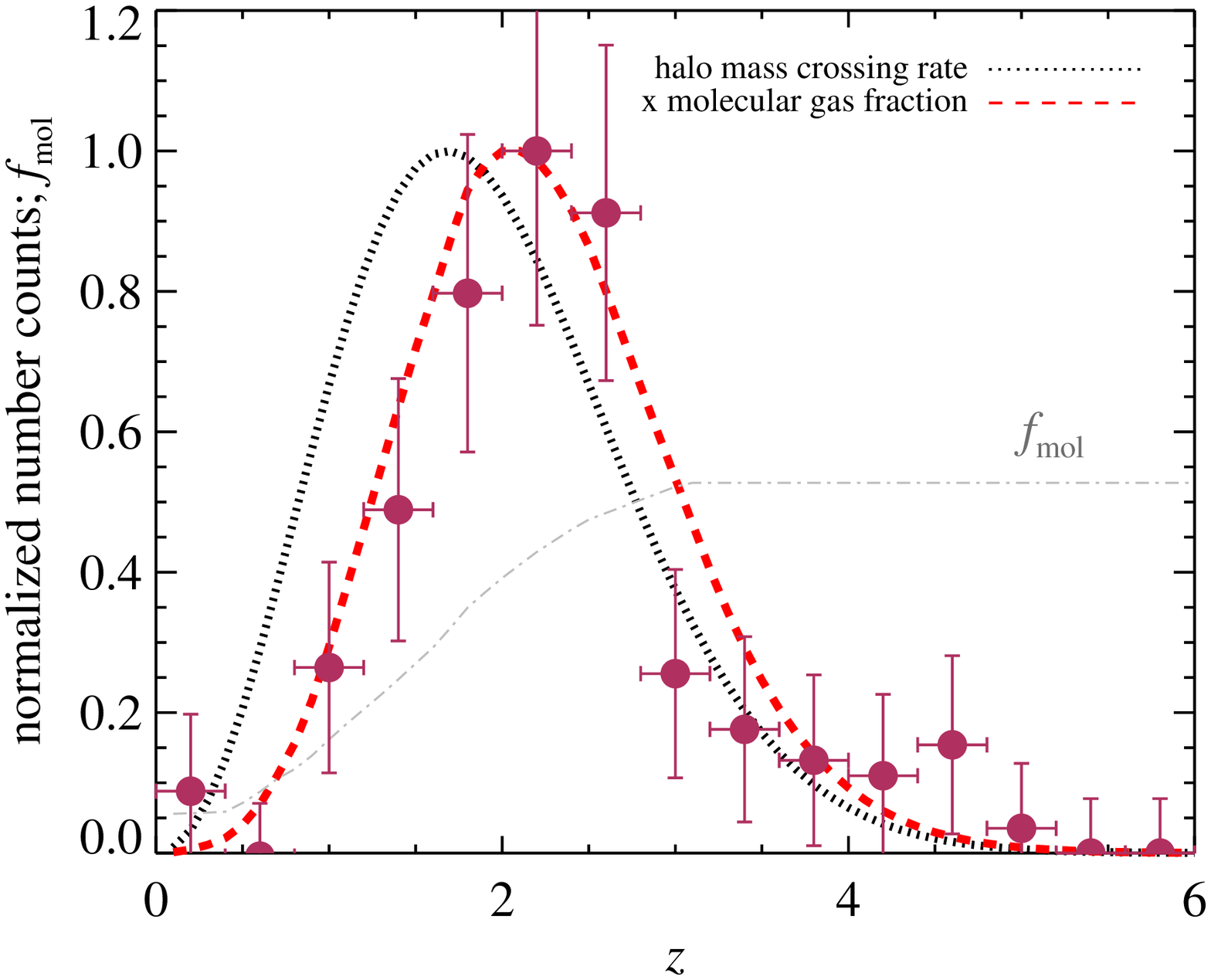}
  \caption{Redshift distribution of LESS SMGs \citep{ward11less},
    compared to the simple models for SMG triggering based on the rate
    at which haloes cross a threshold mass $M_{\rm thresh} =
    6\times10^{12}$ \hmsun\ (see \S~\ref{sec.evol}). The uncertainties
    in the number counts are an approximation of Poisson counting statistics \citep{gehr86}. The
    black dotted line shows the (arbitrarily normalized) number of
    haloes crossing this threshold in each redshift interval
    (Equation~\ref{eqn.thresh}) while the dashed red line shows this
    distribution multiplied by the evolution in the molecular gas
    fraction (Equation \ref{eqn.fmol}), where $f_{\rm mol}$ is taken from the model predictions of \citet{lago11fgas} and is shown by the gray
    dot-dashed line. The remarkable agreement between the second model
    and the observed number counts suggests that the evolution of the
    SMG population can be described simply in terms of two quantities:
    the growth of DM structures and the variation with
    redshift of the molecular gas fraction in galaxies.
    \label{fig.rate}}
\end{figure}

Finally, the observed clustering of SMGs provides insights into the
processes that trigger and (possibly) shut off their rapid star
formation activity. As discussed in \S~\ref{sec.intro}, powerful local
starbursts (i.e. ULIRGs) are predominantly associated with major
mergers and appear to be associated with the fueling of luminous QSOs
as part of an evolutionary sequence \citep[e.g.,][]{sand88}.  However
it is unclear if a similar connection exists between SMGs and high-$z$
QSOs. One robust prediction of any evolutionary picture is that SMGs
and QSOs {\em must} display comparable large-scale clustering, since
the evolutionary timescales are significantly smaller than those for
the growth of DM haloes. At all redshifts, QSOs are found in haloes of
similar mass $\sim$\,a few $\times10^{12}$ \hmsun\ (e.g.,
\citealt{croo05, myer06clust, daan08clust, ross09qsoclust};
Figure~\ref{fig.r0}). The characteristic $M_{\rm halo}$ provides a
strong constraint on models of QSO fueling by the major mergers of
gas-rich galaxies \citep[e.g.,][]{kauf00merge, spri05, hopk06merge},
secular instabilities \citep[e.g.,][]{mo98disk, bowe06gal,
  genz08ifsz2} or accretion of recycled cold gas from evolved stars
\citep{ciot07flare, ciot10flare}, and is similar to the mass at which
galaxy populations transition from star-forming to passive (e.g.,
\citealt{coil08galclust, brow08halo, conr09halo, tink10clust}). The
observed clustering of SMGs at $z\sim 2$ from the present work is
consistent with that for QSOs, as well as highly active obscured
objects including powerful obscured AGN (\citetalias{hick11qsoclust};
\citealt{alle11xclust}) and dust-obscured galaxies
\citep{brod08dogclust}. Thus these may indeed represent different
phases in the same evolutionary sequence, and energy input from the
QSO may be responsible for the rapid quenching of star formation at
the end of the SMG phase \citep[e.g.,][]{dima05qso, spri05} as
discussed in \S~\ref{sec.lifetime}.

A connection with QSOs may imply that triggering of SMGs is also
related (at least indirectly) to the mass of the parent DM
halo. In this case, the evolution of large-scale structure may broadly
explain why the SMG population peaks at $z\sim$\,2.5 and falls at
higher and lower redshifts. In the simplest possible such scenario,
SMG activity is triggered when the halo reaches a certain mass $M_{\rm
  halo} = M_{\rm thresh}$ (see Figure 16 of \citealt{hick09corr} for a
schematic illustration of this picture). In a given volume, the
number of haloes crossing this mass threshold as a function of redshift
is:
\begin{equation}
\frac{{\rm d}N_{\rm thresh}}{{\rm d}z} \propto n_{\rm halo}(M_{\rm
  thresh},z) \dot{M}_{\rm halo}(M_{\rm thresh},z) t_{\rm SMG}
\frac{{\rm d}V}{{\rm d}z},
\label{eqn.thresh}
\end{equation}
where $n_{\rm halo}$ and $\dot{M}_{\rm halo}$ are the number density
\citep[e.g.,][]{tink08halo} and typical growth rate
\citep{fakh10halorate}, respectively, of haloes of mass $M_{\rm
  thresh}$ at redshift $z$, $t_{\rm SMG}$ is the SMG lifetime, and
${\rm d}V/{\rm d}z$ is the differential comoving volume over the
survey area. If an SMG is triggered every time a halo reaches $M_{\rm
  thresh}$, then the observed number density of SMGs will be
proportional to ${\rm d}N_{\rm thresh}/{\rm d}z$.  However, the huge
star formation rates of SMGs require a large reservoir of molecular
gas \citep[e.g.,][]{grev05smgco,tacc06smgco, tacc08smg}, and the
molecular gas fraction increases strongly with redshift
\citep[e.g.,][]{tacc10fgas, geac11fgas, lago11fgas}. This evolution
may explain why the most powerful starbursts at low redshift (ULIRGs)
have lower typical SFRs than $z\sim 2$ SMGs
\citep[e.g.,][]{lefl05irlf, rodi11irlf}. Therefore it may be
reasonable to assume that the number counts of SMGs also depend on
$f_{mol}$, with the simplest possible prescription being:
\begin{equation}
\frac{{\rm d}N_{\rm SMG}}{{\rm d}z} \propto \frac{{\rm d}N_{\rm thresh}}{{\rm d}z} f_{\rm mol}(z).
\label{eqn.fmol}
\end{equation}

In Figure~\ref{fig.rate} we show the observed redshift distribution of
LESS SMGs \citep{ward11less}, compared to the distributions predicted
by Equations (\ref{eqn.thresh}) and (\ref{eqn.fmol}), assuming $M_{\rm
  thresh} = 6\times10^{12}$ \hmsun. For simplicity, the evolution in
$f_{\rm mol}$ is taken from predictions of the {\sc GALFORM} model of
\citet{lago11fgas}, which agrees broadly with observations (see
Figure~2 of \citealt{geac11fgas}) and so provides a simple
parameterisation of the current empirical limits on the molecular gas
fraction in galaxies.  It is clear from Figure~\ref{fig.rate} that
there is remarkable correspondence between our extremely simple
prescription and the observed redshifts of SMGs. Of course this
``model'' does not account for a wide range of possible complications
and the normalisations of the distributions are arbitrary. However,
this exercise clearly demonstrates that if SMGs, like QSOs, are
found in haloes of a characteristic mass, then their observed redshift
distribution may be explained simply by two effects: the cosmological
growth of structure combined with the evolution of the molecular gas
fraction. Thus SMGs likely represent a short-lived but universal phase
in massive galaxy evolution, associated with the transition between
cold gas-rich, star-forming galaxies and passively evolving systems.

\section{Conclusions}

In this paper we measure the cross-correlation between SMGs and
galaxies in the LESS survey of ECDFS, and observe significant
clustering at the $>$\,$4\sigma$ level. We obtain an autocorrelation
length for the SMGs of $r_0 = 7.7^{+1.8}_{-2.3}$ \hmpc, assuming
$\gamma=1.8$. This clustering amplitude corresponds to a
characteristic DM halo mass of $\log{(M_{\rm halo} [h^{-1}\;
  M_{\sun}])}= 12.8^{+0.3}_{-0.5}$. Using this estimate of
$M_{\rm halo}$ and the space density of SMGs, we obtain a typical SMG
lifetime of $t_{\rm SMG} = 110^{+280}_{-80}$ Myr.

The observed clustering indicates that the low-redshift descendants of
typical SMGs are massive ($\sim$\,2--3 $L^*$) elliptical galaxies at the
centers of moderate- to high-mass groups.  This prediction is
consistent with previous suggestions based on the dynamical
\citep{swin06smg} and stellar masses \citep[e.g.,][]{hain11smgmass} of
SMGs, and is also consistent with observations of local massive
ellipticals, which indicate that they formed the bulk of their stars at
$z>2$ and have been largely passive since. The clustering of SMGs is
very similar to that observed for QSOs at the same redshifts,
consistent with evolutionary scenarios in which SMGs and QSOs are
triggered by a common mechanism. Assuming that SMGs, like QSOs, are
transient phenomena that are observed in haloes of similar mass at all
redshifts, the redshift distribution of SMGs can be explained
remarkably well by the combination of the cosmological growth of
structure and the evolution of the molecular gas fraction in galaxies.

This accurate clustering measurement thus provides a valuable
observational constraint on the role of SMGs in the cosmic evolution
of galaxies and large-scale structures. We conclude that SMGs likely
represent a short-lived but universal phase in massive galaxy
evolution that is associated with the rapid growth of black holes as
luminous QSOs, and corresponds to the transition between cold
gas-rich, star-forming galaxies and passively evolving systems.

\section*{Acknowledgments}

We thank the anonymous referee for helpful comments. RCH acknowledges
support through an STFC Postdoctoral Fellowship and AMS from an STFC
Advanced Fellowship. IRS, DMA, ALRD, and JPS acknowledge support from
STFC. IRS acknowledges support through a Leverhulme Research
Fellowship.  DMA is grateful to the Royal Society and the Leverhulme
Trust for their generous support. ADM was generously funded by the
NASA ADAP program under grant NNX08AJ28G. JSD acknowledges the support
of the European Research Council through the award of an Advanced
Grant, and the support of the Royal Society via a Wolfson Research
Merit award. This study is based on observations made with ESO
telescopes at the Paranal and Atacama Observatories under programme
numbers: 171.A-3045, 168.A-0485, 082.A-0890 and 183.A-0666.


\label{lastpage}

\end{document}